\documentclass[doublecol]{epl2}
\usepackage{color}
\usepackage{graphicx}
\usepackage{epsfig}
\usepackage{amsmath, amsthm, amssymb}
\usepackage{enumerate}
\usepackage[normalem]{ulem}
\usepackage{subfig}

\newcommand{\be}{\begin{equation}}
\newcommand{\ee}{\end{equation}}
\newcommand{\bea}{\begin{eqnarray}}
\newcommand{\eea}{\end{eqnarray}}

\title{Improved Upper Bounds on the Asymptotic Growth Velocity of Eden Clusters}
\author{Aanjaneya Kumar and Deepak Dhar}
\institute{  Department of Physics, Indian Institute of Science Education and Research, Pune\\
Dr. Homi Bhabha Road, Pashan, Pune 411008, India}


\abstract{ We consider the asymptotic shape of clusters in the Eden model on a $d$-dimensional hypercubical lattice. We discuss two improvements for the well-known upper bound  to the growth velocity in different directions  by that of the independent branching process (IBP). In the IBP, each cell gives rise to a daughter cell at a neighboring site at a constant rate. In the first improvement, we do not allow such births along the bond connecting the cell to its mother cell. In the second, we iteratively evolve the system by a growth as IBP for a duration $\Delta t$, followed by culling process in which if any cell produced  a descendant within this interval, who occupies the same site as the cell itself, then the descendant is removed.  We  study the improvement on the upper bound on the velocity for different dimensions $d$.   The bounds are asymptotically exact in the large-$d$ limit. But in $d =2$, the improvement over the IBP approximation is only a few percent. 
}

\begin{document}
\maketitle

\section{Introduction}   The first passage percolation problem (FPP) has  a long history. 
Hammersley and Welsh introduced First Passage Percolation as a lattice model of fluid flowing through a random porous medium in 1965 \cite{grow}. Subsequently, it was shown by Richardson and improved upon by Cox-Durrett and Kesten that for this model the number of lattice sites that the fluid can wet grows linearly with time and this cluster of wetted sites, asymptotically converges to (under suitable normalization) a deterministic subset of the lattice called the limit shape \cite{grow_1,grow_2,grow_3}. The problem has attracted a lot of researchers, and a good recent review of the subject may be found in \cite{fpp}.  However, there are many unanswered questions. In particular, the exact solution for the  asymptotic growth velocity  is not known for any regular translationally invariant lattice,  and not much is known about the exact  asymptotic shape of the wetted cluster, beyond its convexity. 

In recent years, there has been a surge in interest in the study of stochastic growth models \cite{grow,grow1,grow2,fpp}. Apart from their many applications, starting from growth of bacterial colonies \cite{bact} to spreading of rumours in a society \cite{rum}, these models have given us a lot of insight into nonequilibrium phenomena by providing us with a platform to study universal behavior \cite{grow1,kpz}. An important question about these processes is understanding the extent of growth of the cluster in different directions, and its fluctuations \cite{fpp}. In this regard, in the context of the Eden model, bounds on velocity of the growing cluster along the axis and the diagonal directions have been obtained earlier \cite{dhar1,kest,35,22}. On a $d$-dimensional hypercubical lattice, the problem becomes easier in the limit of large $d$.  It was shown earlier \cite{dhar2} that the asymptotic velocity along the axis $u_{axis}(d)$ in large dimensions is given by
\begin{equation}
\lim_{d\rightarrow\infty}u_{axis}(d)\frac{\log d}{d}=2
\end{equation}
However, the convergence  to this value is slow. We will give an independent proof of this result below using the fact that the Eden growth process is slower than the independent branching process (IBP). 
For the velocity along the diagonal direction $(1,1,1, ..)$, one can get upper and lower bounds on the velocity, both of which grow as $\sqrt{d}$, with $d$.

The FPP problem is related to the problem of self-avoiding walks, and bounds on the growth constant of self-avoiding walks ( The growth constant  is defined as the limit of $(C_N)^{\frac{1}{N}}$, where $C_N$ is the number of self avoiding walks of $N$ steps). In the 'large $d$' expansion technique,  the growth constant $\mu_d$ of self-avoiding walks  for the $d-$dimensional hypercubical lattice,  is expanded in a asymptotic series of the form $\mu_d =2 d -1 - \frac{1}{2d} +{\mathcal O}(1/d^2)$  \cite{larged}. It is interesting to note that the value $2d$ is the growth constant of fully random walks, and the leading correction to the growth constant comes from disallowing immediate retraversals. We will not need to invoke the connection to self-avoiding walks in this paper, but will instead explore the idea that if we disallow immediate retraversals in IBP, it  would be expected to improve the upper bounds on velocity significantly.
  
As far as the asymptotic shape of the growing cluster is concerned, most studies have been concerning the inequality of the growth velocity along the axes and diagonal. It was first proved by Kesten \cite{kest} that in very high dimensions (greater than about $10^6$), the asymptotic shape of the Eden cluster is not a Euclidean ball. Subsequently, dimensional improvements have been made \cite{35,22} and we now know that the limit shape shows a departure from the Euclidean ball in $d>22$. While some numerical studies of the asymptotic shape have been reported in $d=2$ \cite{2d}, we could not find  any general discussion of the equation of the asymptotic shape, apart from a few simplified models of first passage percolation \cite{shape}.

In this paper, we  first re-derive the already known upper bounds on the growth velocity of Eden clusters on a 
$d$-dimensional hypercubical lattice by studying the Independent Branching Process as an upper bound for the Eden process.  This is then easily extended to get bounds in a general direction and its limit shape. Then, as an improvement over our approximation, we study a birth  process in which any agent gives rise to progeny at neighboring sites at a constant rate, except at the site of its own  parent. By analogy with self-avoiding walks, we expect that this takes care of the leading correction term in the large $d$ expansion of the first passage velocity.  In the second, we iteratively evolve the system by a time-step  process of growth as IBP for a duration $\Delta t$, followed by culling process in which if any cell produced  a descendant within this interval, who occupies the same site as the cell itself, then the descendant is removed. Thus, the finite-time propagator of the branching process is modified so that its value at any site does not exceed one.

\section{The Eden Model}
The Eden Model was first introduced by Murray Eden in 1961 \cite{eden} to investigate the growth of biological cell colonies. Many variants of this model have been studied since then $-$ starting from the model of skin cancer by Williams and Bjerknes \cite{skin} to the SIR (Susceptible-Infected-Recovered) and SIS (Susceptible-Infected-Susceptible) models of epidemics \cite{sirs}. 

We begin by defining the Eden model as an epidemic model. Consider an infection process on a $d-$dimensional hypercubical lattice where each site can either be infected or healthy. We denote the coordinates of each site by $(x_1,x_2,...,x_d)$. At time $t=0$, only the origin $O$=$(0,0,...,0)$ is infected and all other sites are healthy. The evolution is a continuous time Markov process in which an infected site infects each of its healthy neighbours at rate $1$. We consider the process in which an infected site never recovers. This model is equivalent to first passage percolation with exponentially distributed independent passage times. 
In the context of this epidemic process, the question of finding upper bounds to the shape of the growing cluster is natural, and is equivalent to estimating the size of region in which the residing people could be exposed to a particular infection given that the infection started from the origin. We will provide upper bounds to the asymptotic shape of the Eden cluster using the Independent Branching Process (IBP) and its variants. 

\section{The independent branching process}

The independent branching process on the ${Z}^d$ lattice is defined as follows: We consider an infection process in which the number of cells present at a site can be arbitrarily large. Let $n(\vec{R},t)$ denote the number of cells present at the site $\vec{R}$ at time $t$.   At the time $t=0$, there is only one cell present in the system, and it is placed at the origin $\vec{O}$. 
Then, we have $n(\vec{R},t=0)= \delta_{\vec{R},\vec{O}}$.

The time evolution is a continuous-time Markov process. At any time $t$, a cell  gives birth to  a descendant cell, that sits at a nearest neighbor site. Then number of cells at the neighbor increases by one.  Once born, a cell never dies. We assume that the rate at which  a cell gives birth to a daughter cell is $1$ along each bond, independent of the number of cells present at the site, or at neighbours.

In this model, the number of cells present at time $t$ increases exponentially with $t$. Each cell gives birth to a daughter cell at a rate $2d$ per unit time (because each has $2d$ neighbors).  Hence the average number of cells present at time $t$ is  $\exp (2 d t )$, for all $t>0$.  Also, with time, the region occupied by at least one cell, also called the region invaded by the cells, grows with time. The outer boundary of the region invaded by the cells is called the invasion front. We define $u(\vec{\Omega})$ as the velocity of the invasion front in the direction $\vec{\Omega}$ as 
\begin{equation}
u(\vec{\Omega}) = \lim_{t \rightarrow \infty} (1/t) ~\vec{R}_t(\vec{\Omega})
\end{equation}
where $\vec{R}_t(\vec{\Omega})$ is position of  the invasion front in the direction $\vec{\Omega})$ at time $t$. It is easily seen that $u(\vec{\Omega})$ has a non-zero limit, and the fluctuations in $(1/t) ~\vec{R}_t(\vec{\Omega})$ tend to zero as time increases.

In the Eden process(EP), the number of cells at any site is at most 1. It is easily that if we have two configurations $\mathcal{C}$ and $\mathcal{C}'$, where $\mathcal{C}$ evolves according to the rules of EP, and $\mathcal{C}'$ evolves as an IBP, and at any given time $t$,  the number of cells in $\mathcal{C'}$ at any site $\vec{R}$  is greater than or equal to the number at the corresponding site   in   $\mathcal{C}$. Then, this property will be preserved at subsequent times. This implies that the invasion front velocity in IBP provides an upper bound to the invasion front velocity in EP in all directions $\vec{\Omega}$.

It is straight forward to determine the growth velocity in IBP. Let the average number of cells in the IBP at time $t$, at the site $\vec{R}$ be denoted by $\bar{n}(\vec{R},t)$. We use the fact that in IBP, these variables satisfy a linear equation

\begin{equation}
\frac{d}{dt} \bar{n}(\vec{R},t) = \sum_{nn} \bar{n}(\vec{R}',t)
\end{equation}
where the  sum runs over the $2d$ nearest neighbors of $\vec{R}$. This is a linear equation, and is easily solved, by  Fourier transformation.  We define the variables $\tilde{n}(\vec{k},t)$ as
\begin{equation}
\tilde{n}(\vec{k},t) = \sum_{\vec{R}} \bar{n}(\vec{R},t) exp(-i \vec{k}.\vec{R})
\end{equation}

Then, these variables satisfy the equation
\begin{equation}
\frac{d}{dt} \tilde{n}(\vec{k},t) = \lambda(\vec{k}) \tilde{n}(\vec{k},t)
\end{equation}
with $ \lambda(\vec{k}) = 2 \sum_{i=1}^{d}  cos(k_i)$. 

This equation is easily solved, and by inverse Fourier transformation, we get

\begin{equation} 
\bar{n}(\vec{R},t) = \int \frac{d\vec{k}}{(2 \pi)^d} ~\exp( \lambda(\vec{k}) t + i \vec{k}.\vec{R}).
\end{equation}

It is easily seen that for fixed $\vec{r}$, $\bar{n}(\vec{R},t)$ increases as $\exp(2 d t)$. We are interested in the case where as $t$ increases, $\vec{R}$ also becomes bigger with time as  $ \vec{R}= \vec{v}t$.  Then, the integral becomes

\begin{equation} 
\bar{n}(\vec{R},t) = \int \frac{d\vec{k}}{(2 \pi)^d} ~\exp( t \left[ \lambda(\vec{k})  + i \vec{k}.\vec{v} \right]).
\label{eq:barn}
\end{equation}

In the limit of large $t$, this is evaluated easily, using the steepest descent method. The stationary point occurs at a imaginary value of $\vec{k} =i\vec{\kappa}$, given by
\begin{equation}
\kappa_j = \sinh^{-1}( v_j/2).
\end{equation}
We define the large deviation function $F(\vec{v})$ by the condition that for large $t$, 
\begin{equation}
\bar{n}(\vec{v}t,t) \sim \exp \left[ t F(\vec{v})\right]
\end{equation}
with
\begin{equation}
F(\vec{v}) = \sum_{i=1}^{d} \left[ 2 \sqrt{ 1 + {v_i}^2/4} - v_i \sinh^{-1} (v_i/2) \right]
\end{equation}

We note that $F$ is a decreasing function of its argument. At $\vec{V}=0$, it has a value $2d$. And for large $|v|$, it varies as $- \sum_i |v_i| \log |v_i|)$. 

At the cluster boundary, $\bar{n}$ is of $\mathcal{O}(1)$. So, the boundary of the cluster, scaled by $t$, is  given by equating the growth rate of $\bar{n}$ to zero. Thus, we get that the scaled boundary of the cluster in the IBP  is given by 
\begin{equation}
\sum_{i=1}^{d} \left[ 2 \sqrt{ 1 + {v_i}^2/4} - v_i \sinh^{-1} (v_i/2) \right]=0.
\label{eq:shape_IBP}
\end{equation}

Note that from Eq.(\ref{eq:barn}), $\bar{n}(\vec{R},t)$ is also equal to the number of walkers expected at $\vec{R}$ at time $t$, if $\exp(2 d t)$ walkers are released at the origin at time $t=0$, and perform independent random walks. 

As a check, we see that along the diagonal direction $(1,1,1,1..)$, we set $v_i = v^*$, for all $i$. Then, for all $d$,  we get $V^*$ is the solution of the equation
\begin{equation}
 2 \sqrt{ 1 + {v^*}^2/4} = v^* \sinh^{-1} (v^*/2)
\end{equation}
which gives $v^* \approx 3.01776$. This gives  the well-known upper bound to the speed along the diagonal in $d$ dimensions (measured in Euclidean norm) as 
\begin{equation}
v_{diag, EP} \leq  v_{diag, IBP} = 3.01776 \sqrt{d}. 
\end{equation}
\begin{figure}
\subfloat[]{\includegraphics[scale=0.47]{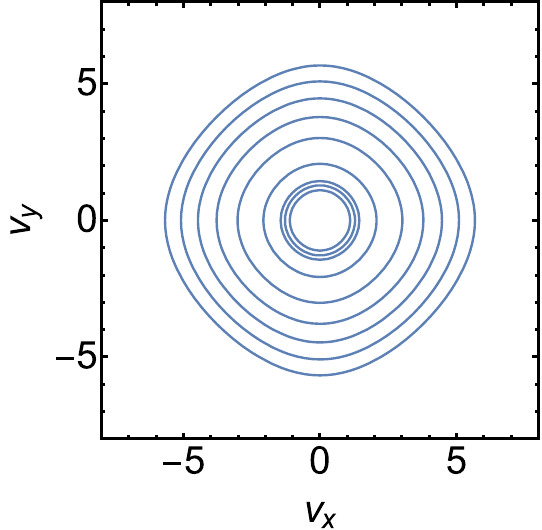}}
\subfloat[]{\includegraphics[scale=0.46]{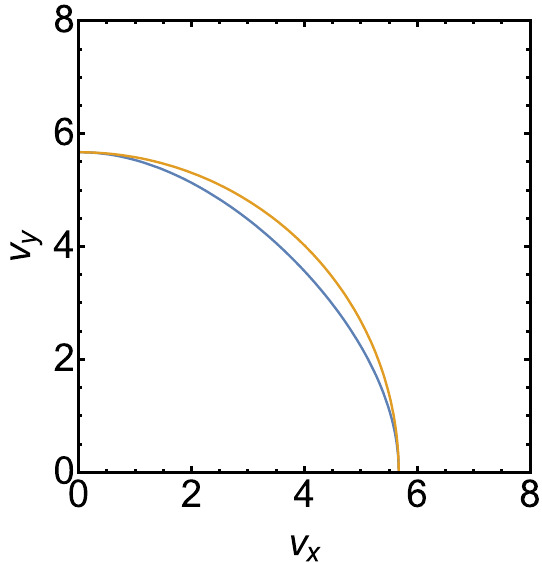}}
\caption{(a) A contour plot for the independent branching process in $d=3$ for $z=0,1,2,3,4,5,5.5,5.65$ starting with the outermost curve. (b) A demonstration of the departure from the spherical ball. The plot in blue is a numerical plot of Eq(10) and orange is an arc of a circle with the radius $5.67$.}
\end{figure}

One can also verify that this agrees with the already known results about  the asymptotic velocity along one of the axes in the IBP. In this case, we set
$\vec{v}=(V_{axis,IBP},0,0,0..)$.  The corresponding equation becomes
\begin{equation}
2 \sqrt{ 1 + \frac{{V_{axis,IBP}^2}}{4}} + 2d -2  = V_{axis,IBP} \sinh^{-1} (\frac{V_{axis,IBP}}{2})
\end{equation}
This is easily solved, and gives $V_{axis,IBP} = 4.4668, 5.67295,6.75371$ and $7.75405$, for $d=2,3,4$ and $5$ respectively. For large $d$, $V_{axis}$ varies as $2d/log(d)$.  

\begin{figure}
    \centering
    \includegraphics[scale=0.4]{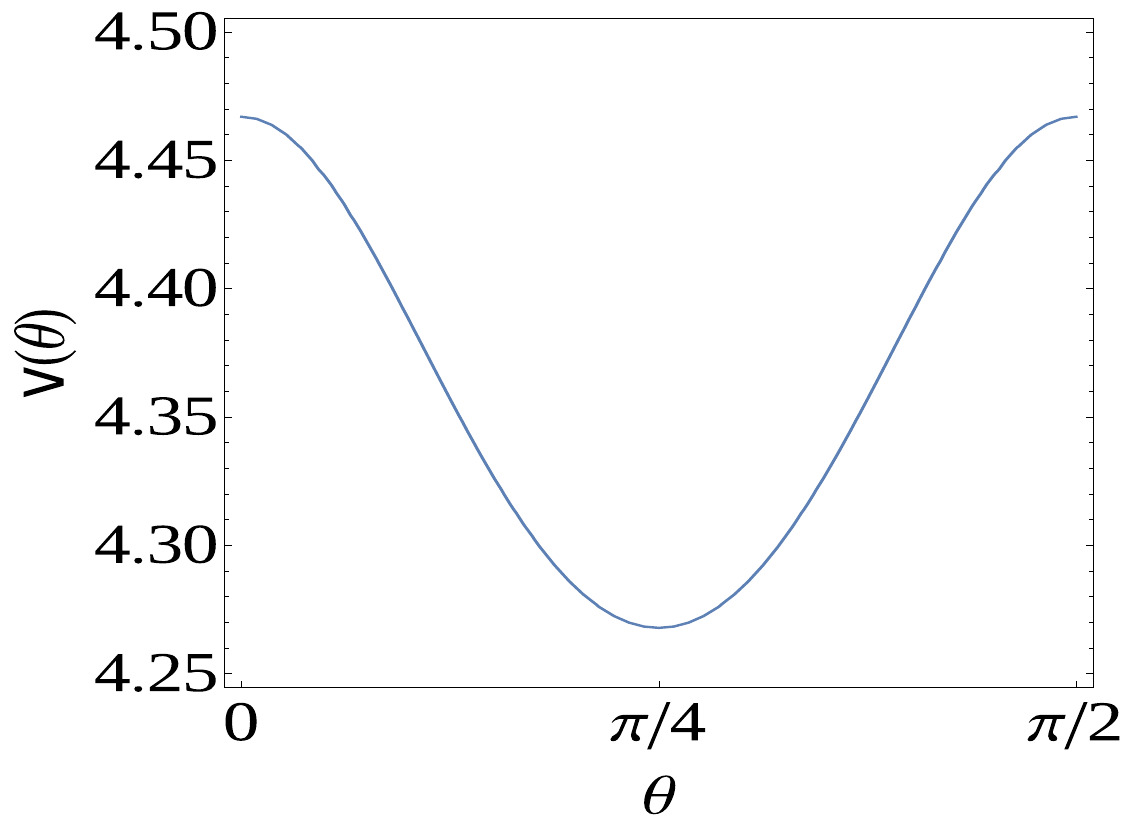}
    \caption{A plot of velocity $v$ as a function of direction $\theta$ for the IBP in $d=2$. Note that the velocity is maximum along the axis ($\theta=n\pi/2$ for integer $n$) and minimum along the diagonal direction ($\theta=(2n+1)\pi/4$).}
    \label{fig:my_label}
\end{figure}

These results about velocity along the axes, or along the main diagonal have been known for a long time.  However, we did not find an explicit discussion of the  asymptotic shape of the cluster for the IBP in a general direction in the  literature.  In Fig. 1, we show the asymptotic shape in $3-d$ calculated numerically 
using Eq.(\ref{eq:shape_IBP}. 
In the EP, Alm and Deijfen found that the cluster shape is not exactly circular, with the Euclidean speed along the diagonal and along the axis  being $2.4420$ and $2.4742$ respectively.  Thus the speed along the diagonal is smaller by about $1.3\%$.  For the IBP, we found these to be $4.26775$ and $4.4668$, with the diagonal speed being smaller that that along the axis by a bigger amount(about $4.5\%$).

\section{ The First Variation of IBP (${\rm IBP}_1$)}

We will now define a variation that is a bit more complicated than the IBP defined above. This is also a continuous-time Markovian evolution independent  branching process. Here, we consider the process on a hyper-cubical lattice in $d$ dimensions. The number of cells at any site can be a non-negative integer. As before, We start with a single cell at the origin at $t=0$, with the rest of the lattice empty. However, We note that all cells, other than the original `eve'-cell have a mother cell. The evolution rule is still Markovian. Each cell gives rise to a daughter cell along each bond at rate $1$, independent of the state of the other sites, {\em except along the bond that connects it to its mother cell.} Thus, such a cell will have $(2d -1)$ bonds along which it can  The `eve'-cell gives rise to a child along each of the $2d$ bonds connected to it, at rate $1$. We call this process the  ${\rm IBP}_1$. 

For the IBP$_1$ process also, we can write a closed set of coupled linear evolution equations for the average number of cells at site $\vec{R}$ at time $t$. But we need to define $2d$ variables at each site. Let $\bar{n}(\vec{R}, t, {e_\alpha} )$ denote the average number of cells residing at the site $\vec{R}$ at time $t$, whose mother cell is along the bond $\bf{e}_{\alpha}$. 
Here $\alpha$ takes $2 d$ possible values $\pm 1, \pm 2, \ldots ,\pm d$, and $e_1$ is the unit vector along coordinate $x_1$, and $e_{-1} = -e_1$. 
Then, the variables $\bar{n}(\vec{R},t,e_\alpha)$ evolve according to the equations
\begin{equation}
\frac{d}{dt} \bar{n}(\vec{R},t, e_\alpha) = \sum_{\alpha' \neq -\alpha} n(\vec{R} + 
e_\alpha, t, e_{\alpha'})
\end{equation}

Again, we define the Fourier transform variables of $\bar{n}(\vec{R},t,e_\alpha)$ as 
$\tilde{n}(\vec{k},t,e_\alpha)$ as 
\begin{equation}
\tilde{n}(\vec{k},t,e_\alpha)=\sum_{\vec{R}} \exp( -i \vec{k}.\vec{R}) \bar{n}(\vec{R},t, e_\alpha)
\end{equation}
Then, the equations for different $\vec{k}$ decouple, and the infinite set coupled equations reduces to that of $2d$ coupled variables $\tilde{n}(\vec{k},t,e_\alpha)$, for the $2d$ values of $\alpha$ for fixed $\vec{k}$.  These are easily seen to be
\begin{eqnarray}
\frac{d}{dt} \tilde{n}(\vec{k},t,e_\alpha) = \exp(i k_\alpha)\left[ S(\vec{k})- \tilde{n}(\vec{k},t,-e_\alpha)\right]\\
\frac{d}{dt} \tilde{n}(\vec{k},t,-e_\alpha) = \exp(-i k_\alpha)\left[ S(\vec{k})- \tilde{n}(\vec{k},t,e_\alpha)\right]
\end{eqnarray}
where $\alpha= 1,2..d$ and $S(\vec{k}) = \sum_{\beta=1}^{d} [ f_{\beta} + f_{-\beta}]$ with $f_{\beta}$ and $f_{-\beta}$ represent $\tilde{n}(\vec{k},t,e_\beta)$ and $\tilde{n}(\vec{k},t,-e_\beta)$ respectively. 
This may be written as
\begin{equation}
\frac{d}{dt}\tilde{n} ( \vec{k},t,\alpha) = \sum_{\alpha'} M_{\alpha, \alpha'}(\vec{k}) ~\tilde{n}(\vec{k},t, \alpha') 
\end{equation}

The eigenvalues for this $2d \times 2d$ matrix for a fixed $\vec{k}$-block  are easily determined.  Explicitly, the matrix elements are
\begin{eqnarray}
M_{\alpha,\beta} &=& \exp( i k_{\alpha}), ~~{\rm ~if } \beta \neq -\alpha;\\
     &=& 0,{\rm ~ if } \alpha = -\beta.
\end{eqnarray}
Here the indices $\alpha$ and $\beta$ take values $(\pm1,\pm2,\pm3 ..\pm d)$
For the eigenvalue $\lambda$, let the eigenvector of the matrix be $f_{\alpha}$, we have, for $\alpha = 1$ to $d$
\begin{eqnarray}
\lambda f_{\alpha}  = e^{i k_{\alpha}} [ S(\vec{k}) - f_{-\alpha}]\\
\lambda f_{-\alpha}  = e^{-i k_{\alpha}} [ S(\vec{k}) - f_{\alpha}]
\end{eqnarray}
where 
\begin{equation}
S(\vec{k}) = \sum_{i=1}^{d} = [ f_i + f_{-i}]. 
\label{eq:S}
\end{equation}
~\\
We try to solve the coupled equations for $f_{\alpha}$ and $f_{-\alpha}$ in terms of $S(\vec{k})$. 
Two cases arise. If $\lambda^2 \neq 1$, then, we can solve these equations to give
\begin{equation}
f_{\alpha}= \frac{e^{i k_\alpha}\lambda -1}{\lambda^2 -1} S(\vec{k}); f_{-\alpha}= \frac{e^{-i k_\alpha}\lambda -1}{\lambda^2 -1} S(\vec{k})
\end{equation}
Then, the consistency condition  \ref{eq:S} then becomes
\begin{equation}
\lambda^2 - 1 +2d = 2 \lambda \left[\sum_{i=1}^{d} \cos k_i \right]
\end{equation}
This is a quadratic equation in $\lambda$, and has two roots. We denote the larger one by 
$\lambda_{max}$. this gives
\begin{equation}
\lambda_{max} = \Lambda +\sqrt{\Lambda^2 -2 d +1}.
\end{equation}
where we have used the abbreviation $\Lambda= \sum_{i=1}^d \cos (k_i)$.
~\\
On the other hand, if $\lambda = \pm 1$, then to get a non-zero solution, we must have $S(\vec{k})=0$, and $f_{\alpha} = -\lambda e^{i k_\alpha} f(-\alpha)$, for $\alpha= 1$ to $d$.

Writing  $\vec{R}=\vec{u}t$ allows us to write
\begin{equation}
\bar{n}(\vec{u}t,t)=  \int \frac{d\vec{k}}{(2 \pi)^d} \exp \left[ t\left(i\vec{k}.\vec{u}+\Lambda +\sqrt{\Lambda^2 -2 d +1} \right) \right]
\end{equation}
The above integral can be easily evaluated in the long time limit using the method of steepest descent giving
\begin{equation}
    iu_j - \sin{k_j}\left[ \frac{\sqrt{\Lambda^2 -2 d +1} + \Lambda}{\sqrt{\Lambda^2 -2 d +1}}\right] = 0
\end{equation}
The stationary point occurs at a imaginary value of $\vec{k} =i\vec{\kappa}$ which upon substitution gives
\begin{equation}
    \kappa_j=\sinh^{-1}{\frac{u_j}{\beta}}
\end{equation}

where 
\begin{equation}
  \beta =  \frac{\sqrt{\Lambda^2 -2 d +1} + \Lambda}{\sqrt{\Lambda^2 -2 d +1}}
\end{equation}

This gives us 
\begin{equation}
  \Lambda = \sum_{j=1}^{d} \sqrt{1+(\frac{u_j}{\beta})^2}
\end{equation}

We can self-consistently determine $\Lambda$ and $\beta$ for given $\vec{u}$ from Eqs.(31,32). Now we can write 
\begin{equation}
\bar{n}(\vec{u}t,t) \sim \exp \left[ t (\Lambda +\sqrt{\Lambda^2 -2 d +1}-\sum_{j=1}^{d} u_j\sinh^{-1}{\frac{u_j}{\beta}})\right]
\end{equation}
Let $\Lambda^*(\vec u)$, $\beta^*(\vec u)$ be solutions of Eqs.$(31,32)$. Then the equation of the surafce is obtained to be
\begin{equation}
     \sum_{j=1}^{d}\left[ u_j\sinh^{-1}{\frac{u_j}{\beta^*(\vec u)}}\right]= \Lambda^*(\vec u) +\sqrt{\left(\Lambda^*(\vec u)\right)^2 -2 d +1}.
\end{equation}

By a straightforward numerical calculation, we find that in IBP$_1$, the velocity along axis is $4.3466$, $5.3533$, $6.4485$ and $7.4602$ while along the diagonal direction is $3.9770$, $4.9772$, $5.8150$ and $6.5485$ for $d=2$, $3$, $4$ and $5$ respectively.

\section{The Second Variation of IBP (IBP$_2$)}

We define the second variation of the IBP as follows: we start with a single eve cell at the origin at $t=0$, with the rest of the lattice empty. We break the time evolution of the process in time intervals of $\Delta$. In the time interval   $z\Delta   \leq t \leq (z+1) \Delta $ where $z$ is nonnegative integer, the system evolves according to the IBP rules. At times $z \Delta$, we introduce a culling process: If any cell present at time $z \Delta$  at any site $\vec{R}$,  generates a descendant ( this is second-order  descendant, or higher order descendant) within the subsequent interval $\Delta$, and this descendant occupies the same site $\vec{R}$, then the descendant is removed.  
 This modified evolution + culling process is captured by the modified propagator $G'(R,\Delta)$. This new propagator $G'$ is obtained from the IBP propagator $G_0$ by  a process of clipping: we write 
\begin{equation}
G'(R,\Delta)= [ G_0(R,\Delta)]_{clipped}
\end{equation}
where the clipping process on  a function is here defined as replacing its value by $1$ if it is greater than $1$, and leaving it unchanged, if the value is $ \leq 1$.  In our calculations, we  work with low enough values of $\Delta$ so that the propagator $G_0$ at all sites other than the origin is less than $1$ at time $\Delta$. Then, the propagator for this variation is given by
\begin{equation}
G'(\overrightarrow{R},\Delta)=G_{0}(\overrightarrow{R},\Delta)+\left[1-G_{0}(0,\Delta)\right]\delta_{\overrightarrow{R},0},
\end{equation}

Where we have 
In Fourier space, 

\begin{equation}
\widetilde{G'}(\overrightarrow{k},\Delta)=e^{2\Delta\sum\cos k_{i}}+1-I_{0}^{d}(2\Delta)
\end{equation}

To evolve this system up to time $t=n\Delta$, it is immediately noted that the propagator needs to be applied iteratively and the result takes the form
\begin{equation}
\widetilde{G'}(\overrightarrow{k},n\Delta)=\widetilde{G'}(\overrightarrow{k},\Delta)^{n}
\end{equation}


Upon taking an inverse Fourier transform and substituting for $\overrightarrow{R}=\overrightarrow{u}t$, we get 

\begin{equation}
G(\overrightarrow{u}t,t)=\int\frac{\overrightarrow{dk}}{(2\pi)^{d}}e^{t\left(\frac{1}{\Delta}\log\left(e^{2\Delta\sum\cos k_{j}}+1-I_{0}^{d}(2\Delta)\right)+i\overrightarrow{k}.\overrightarrow{u}\right)}
\end{equation}

The above integral can be easily evaluated in the long time limit using the method of steepest descent giving
\begin{equation}
iu_{j}-\frac{2\sin k_{j}e^{2\Delta\sum\cos k_{i}}}{e^{2\Delta\sum\cos k_{j}}+1-I_{0}^{d}(2\Delta)}=0
\end{equation}

The stationary point occurs at a imaginary value of $\vec{k} =i\vec{\kappa}$ which upon substitution gives

\begin{equation}
u_{j}=\frac{2}{\Gamma}\sinh\kappa_{j}
\end{equation}
where

\begin{equation}
\Gamma=\frac{e^{2\Delta\sum\cosh\kappa_{j}}-I_{0}^{d}(2\Delta)+1}{e^{2\Delta\sum\cosh\kappa_{j}}}
\end{equation}

We can eliminate $\kappa$ from the above equation. For notational convenience, we define 
\begin{equation}
A=2 \Delta \sum_{j=1}^{d} \cosh \kappa_j=2\Delta\sum_{j=1}^{d}\sqrt{1+\left(\frac{\Gamma}{2}u_{j}\right)^{2}}.
\end{equation}
Then Eq.($42$) can be rewritten as




\begin{equation}
\ln\left(1-\Gamma\right)=\ln\left(I_{0}^{d}(2\Delta)-1\right)-A
\end{equation}

Let $A^*(\vec u)$ and $\Gamma^*(\vec u)$ be the solution to Eqs.($43,44$) for a given $\vec{u}$. Then the equation of the shape of IBP$_2$ cluster is determined to be
\begin{eqnarray}
\nonumber \sum_{j=1}^{d} \left[ u_j\sinh^{-1}\frac{\Gamma^*(\vec u)}{2}u_{j}-2\sqrt{1+\left(\frac{u_j\Gamma^*(\vec{u})}{2}\right)^2} \right]\\ =\frac{\log\Gamma^*(\vec u)}{\Delta} 
\end{eqnarray}


It is clear that if $\Delta$ is close to zero, then the capping is ineffective as for small $\Delta$, evolution is effectively already capped. For very large $\Delta$, this method would be ineffective as only the origin would be capped and the evolution of the IBP$_2$ would be very much like the IBP. It is also easily verified if we take the $\Delta \to 0$ and the $\Delta \to \infty$ limits, we recover Eq.(11) from Eq.(45). We can choose the optimum value of $\Delta$ that gives the best bound. 

In IBP$_2$, the velocity along axis is evaluated to be $4.1134$, $5.3826$, $6.4927$ and $7.5107$ while along the diagonal direction, it is $3.9653$, $5.0019$, $5.8492$ and $6.5838$ for $d=2$, $3$, $4$ and $5$ respectively.

\section{Summary and discussion}
In this paper, we developed improved upper bounds to the asymptotic shape of Eden clusters. We  first  found the exact equation that gives  the exact asymptotic shape of the IBP cluster. We found that even in the IBP cluster, a departure from the circular shape of cluster is seen as pointed out by Alm and Deijfen for the Eden cluster. Then we improved upon the bounds by considering two independent modifications to the IBP - one in which each cell independently gives rise to daughter cells at neighbouring sites except along the bond that connects it to its mother cell and the other, in which we iteratively evolve the system and in each iteration, impose the condition at a non-empty site, no more cells can be added due to the descendants of the cells present at that site.  Contrary to our initial expectation,  we found that both these modifications   improve  of the upper bound only by a few percent, even for small dimensions like 2  to 5. 

Of more interest is the structure of the equation that describes the dependence of the velocity on direction. We note that the equation that describes the direction dependent velocity in the IBP is eq.(11), which is a condition of the form
\begin{equation}
\sum_{i=1}^{d} g(u_i) = {\rm constant}.
\label{eq:shape}
\end{equation}
Here $g(u)$ is a convex function of its argument, and the equation has explicitly the permutation symmetry over directions. For example, if we  $g(u)$ was of the form $g(u) = {\rm constant} |u|^{a}$, with $ 1<a <2$, we get a set of shapes that interpolate between d-dimensional sphere and d-dimensional diamond-shape. 

 From the explicit form of the function, we see that  $g(\epsilon) \sim \epsilon^2$, for small $\epsilon$.   This implies that if we look at the a direction only slightly away from axes, at a small angle $\theta$ from the axis, we get the velocity in this direction depends only on $\theta$, and is given by $V_{axis,IBP} - a \theta^2$, where $a$ is some constant.  Of course, this is shown only for the IBP, which  only gives an upper bound the the Eden cluster size. 

It it an interesting question to see if the true Eden cluster shape is also described  by an equation of the form of Eq.(46), with a different  convex function $g$.   
The structure of shape in the IBP$_2$ bound is  more complicated. That can only written as  a couple of equations of the form 
\begin{equation}
\nonumber
\sum_{i} g( u_i, \Lambda)= {\rm constant},~~\sum_{i} h( u_i, \Lambda) ={\rm constant}.
\end{equation}
where $g(u, \Lambda)$ and $h(u,\Lambda)$ are convex functions of the argument $u$, and $\Lambda$ is a parameter, which has to be eliminated from the two equations, to generate an equation only amongst the variables $\{ u_{i} \}$. 
Of course, higher order approximations may give rise to coupled set of equations, symmetric in $\{u_i\}$, but with more parameters. This seems like an interesting question to explore further.

This article is dedicated to Joel Lebowitz  for his gentle mentoring  of  many  generations of scientists as the Chief Editor of Journal of Statistical Physics for 40 years.

\bibliographystyle{eplbib}

 \end{document}